\documentclass[12pt]{article}
\usepackage{amsmath,amssymb}
\renewcommand{\a}{\alpha}
\renewcommand{\b}{\beta}
\newcommand{\bs}{\bigskip}

\newcommand{\D}{\Delta}

\renewcommand{\i}{\infty}

\renewcommand{\L}{\Lambda}

\renewcommand{\ni}{\noindent}
\newcommand{\p}{\partial}
\renewcommand{\th}{\theta}
\renewcommand{\t}{turbulen}
\newcommand{\pg}{pressure-gradient }
\newcommand{\ppg}{pressure gradient}
\newcommand{\Rn}{Reynolds number }
\newcommand{\ep}{experimental }
\newcommand{\bl}{boundary layer}
\renewcommand{\ss}{self-similar }
\newcommand{\st}{structure}
\newcommand{\RL}{Re_{\Lambda}}

\begin{document}

\begin{center}
{\large\bf A  Model of a
{\large\bf Turbulent Boundary Layer}\smallskip\\With a Non-Zero Pressure
Gradient}

\vspace{.5 truein}

{\sc G. I. Barenblatt,${}^*$ \ A. J. Chorin${}^*$\\and V. M.
Prostokishin${}^\dagger$}

\vspace{.33 truein}
${}^*$Department of Mathematics and\\
Lawrence Berkeley National Laboratory,\\
University of California,
Berkeley, California 94720--3840;\\
${}^\dagger$P. P. Shirshov Institute of Oceanology,\\ Russian Academy of
Sciences,\\ 36 Nakhimov Prospect, Moscow 117218 Russia

\vspace{.5 truein} \ni {\bf Key Words:} \t t boundary layer, \ \ppg\!, \
self-similarity, \ scaling laws.

\bs Contributed by G.~I.~Barenblatt
\end{center}

\bs\noindent
{\bf Abstract.} {\footnotesize\bf
According to a model of the turbulent boundary layer proposed by the authors,
in the absence of external turbulence
the intermediate region between the viscous sublayer and the external flow
consists of two sharply separated self-similar structures. The velocity
distribution in these structures is described by two different scaling laws.
The mean velocity $u$ in the region  adjacent to the viscous sublayer is
  described by the previously obtained Reynolds-number-dependent scaling law
$\phi = u/u_*=A\eta^{\alpha}$, \ $A=\frac{1}{\sqrt{3}}\,\ln Re_{\Lambda}+
\frac 52$, \ $\alpha=\frac{3}{2\ln Re_{\Lambda}}$,
  \ $\eta = u_* y/\nu$. (Here $u_*$
is the dynamic or friction velocity, $y$ is the distance from the wall,
$\nu$ the kinematic viscosity of the fluid, and the Reynolds number
$Re_{\Lambda}$ is well defined by the data) In the region 
adjacent to the external flow the scaling law is different: $\phi=
B\eta^{\beta}$. The power $\beta$  for zero-pressure-gradient boundary layers
was found by processing various experimental data  and is close (with some
scatter) to 0.2.

We show here that
for non-zero-pressure-gradient boundary layers,  the power $\beta$ is larger
than 0.2 in the case of adverse pressure gradient and less than 0.2 for
favourable
pressure gradient. Similarity analysis suggests that both the coefficient $B$
and the power $\beta$ depend on $Re_{\Lambda}$ and on a new dimensionless
parameter $P$ proportional to the pressure gradient. Recent experimental data of
Perry, Maru\v{s}i\'c and Jones (1)-(4) were analyzed and the results are in
agreement with the model we propose.}

\setlength{\baselineskip}{16pt}

\section{Introduction}
The model of the \t t \bl \ at large \Rn  proposed by Clauser (5) and Coles (6)
is widely accepted and used. This model is based on
the assumpation that the transition from the wall region described by the
  Karman-Prandtl (7)-(8) universal logarithmic law to the external
flow is smooth. On the basis of our analysis of \ep data published  
over
the last 30 years we arrived  at a different model (see (9)-(11)). According to
our model, if the intensity of \t ce in the external flow is low, then the
intermediate region between the viscous sublayer and the external flow
consists of two \ss \st s separated by a sharp boundary. In
particular,  when the the bilogarithmic coordinates
$\lg\phi, \ \lg\eta$ are used, where $\phi =u/u_*$, $\eta = u_* y/\nu$,
the mean velocity profile in the intermediate region has a characteristic
form of a broken line (`chevron') (Figure 1). In part I of the intermediate
region the scaling law for the mean velocity distribution takes the form:
$$
\frac{u}{u_*} \ = \ A\left(\frac{u_*y}{\nu}\right)^\a \ .
\eqno{[1]}
$$
In part II one finds a different scaling law:
$$
\frac{u}{u_*} \ =  B  \left(\frac{u_*y}{\nu}\right)^\b \ .
\eqno{[2]}
$$
The constants $A,\a,B,\b$ can be determined  with
sufficient accuracy by processing the \ep data. According to our model 
the 
expressions for $A$ and $\a$ are identical to those in smooth pipes once 
the Reynolds number is defined correctly:

$$
A \ = \ \frac{1}{\sqrt{3}} \ln Re_{\L}+{\textstyle{\frac 52}} \ ,
\qquad \a \ = \ \frac{3}{2\ln Re_{\L}} \ .
\eqno{[3]}
$$
Here $Re_{\Lambda}$ is an effective \Rn for \t t \bl, different from the
usual \Rn
$Re_{\th}$ based on momentum thickness which is arbitrarily though  
widely used in \t t \bl \ studies. The test of the validity
of our model is the closeness of two values of $\ln\RL$, $\ln Re_1$
and $\ln Re_2$, obtained by solving separately the two equations [3] with parameters $A$
and $\a$ obtained from \ep data. Differences of less
than  2--3\% were obtained in
all cases (see (10),(11) for previous data processing), and therefore we proposed to take
$\ln\RL$ as half the sum $\frac 12(\ln Re_1+\ln Re_2)$.

In region II the power $\b$ in the scaling law [2] for the
zero-\pg \bl \ was found to be close to 0.2 (with some scatter).
In cases of non-zero-\pg \bl s the values of $\b$ were found to be
significantly different from 0.2. In the present Note we perform the
similarity analysis for non-zero-\pg \t t \bl . We find that both the 
coefficient $B$ and the power $\b$ depend on $\RL$ and on a new similarity
parameter $P=\nu\p_x p/\rho u^3_*$. We compare the results of this analysis with high quality
\ep data by Maru\v{s}i\'c and Perry(1), Maru\v{s}i\'c (2), Jones,
Maru\v{s}i\'c and Perry (3), Jones (4), and come to the instructive 
conclusions.

\section{The model and the similarity\\analysis}

According to our model the \t t \bl \ at large Reynolds numbers consists of two
  separate layers I and II. The \st \ of the vorticity fields in the two layers is
different although both are \ss. In layer I the vortical \st \ is the one 
common to all developed wall-bounded shear flows and the mean flow
velocity is described by relations [1] and [3]. In these relations
$\RL=U\L/\nu$, where $\L$ is a characteristic length (12) close to 1.6
of the height of layer I.

The influence of viscosity is transmitted to the main body of the flow
via streaks separating from the viscous sublayer.\footnote{We note that
this mechanism for the molecular viscosity affecting 
the main body of the flow was proposed by L.~Prandtl in his discussion of
  Th.~von K\'arm\'an's lecture
(8). It is rather astonishing that this idea was never repeated in Prandtl's
subsequent publications.} The remaining part of the intermediate region of the
\bl
\ is occupied by layer II where the relation [2] holds.  It is well known  (see
in particular instructive photographs in Van Dyke's {\it Album of Fluid
Motions} (13)) that the upper boundary of the \bl \ is covered with statistical
regularity by large scale `humps' and  that the upper layer is influenced by the
external flow via the form drag of these humps as well as by the shear stress.
We have shown in earlier work that the mean velocity profile is affected by the intermittency
of the turbulence, and as the humps affect intermittency it is natural to see two different scaling
regions. On the basis of these considerations we have 
to determine a set of parameters
that determine 
the coefficient
$B$ and the power $\b$ in [2]. One of these parameters must be 
the effective
\Rn
$\RL$ which determines the flow structure in the layer I and is affected in its turn 
by the viscous sublayer and by layer II. The
following parameters should also influence the flow in the upper layer:
the \ppg $\p_x p$ ($x$ is the longitudinal coordinate reckoned along the
plate; its origin is immaterial), the dynamic (friction) velocity $u_*$, and
the fluid's kinematic viscosity $\nu$ and density $\rho$.
The dimensions of governing parameters are as follows
$$
[\p_x p]=\frac{M}{L^2T^2} \ , \quad
[u_*] =\frac{L}{T} \ ,\quad
[\nu]=\frac{L^2}{T} \ ,\quad
[\rho]=\frac{M}{L^3}\ .  \eqno{[4]}
$$
The first three have independent dimensions so that only 
one dimensionless governing parameter can be formed:
$$
P \ = \  \frac{\nu\p_x p}{\rho u^3_*}  \eqno{[5]}
$$

Thus
the parameters $B$ and $\b$ should depend on two the
parameters $\RL$ and $P$:
$$
B \ = \ B(\RL,P) \ , \qquad
\b \ = \ \b(\RL,P) \ .
  \eqno{[6]}
$$

\section{Comparison with \ep data}

The data for non-zero-\pg flows
are substantially less numerous than data for zero-\pg flows, and do not allow us yet to construct
surfaces $B(\RL,P)$, $\b(\RL,P)$. However the high quality data obtained by
Maru\v{s}i\'c and Perry [(1), recently brought to completion via the internet]
and Jones,   Maru\v{s}i\'c and Perry [(2),
also completed on  the internet] allowed us to
come to some instructive conclusions. The experiments
of Maru\v{s}i\'c and Perry (1) were performed for
two external flow velocities $U$: 10 m/s and 30 m/s.
The experiments of
  Jones, Maru\v{s}i\'c and Perry (3) were performed for
three external flow velocities $U:$ 5 m/s, 7.5 m/s, and 10 m/s. The results of
the 
processing of the \ep data are presented in Table 1.
Here \ {\bf x} \ and $Re_{\theta}$ are given by the authors
of the experiments, and $\Delta=2|\ln Re_1-\ln Re_2|/(\ln Re_1+\ln Re_2)$.

\begin{center}{\bf Table 1}\end{center}
\vspace{-0.125in}
\begin{tabular}{rrrrrrrrrr} \\
{\bf x, m} & ${\boldmath{Re_{\th}}}$ & ${\boldmath\a}$ &
{\bf A} & ${\boldmath\b}$ & {\bf B}  & ${\boldmath{\ln Re_1}}$ &
  ${\boldmath{\ln Re_2}}$ & ${\boldmath{\ln\RL}}$ & ${\boldmath{\D}}$ \\
{\bf I.Maru\v{s}i\'c data} \\
$U=10$ m/s \\
1.20	&2,206	&0.143&	8.53	&0.203&	6.18	&10.44	&10.51&	10.48&	0.7 \\
1.80	&3,153&	0.150	&8.30&	0.227&5.45	&10.05&	10.03&	10.04&	0.2 \\
2.24	&4,155&	0.156	&8.15&0.269	&4.34&	9.79	&9.88	&9.84	&0.9 \\
2.64	&5,395&	0.171	&7.54&	0.345	&2.87&8.73&	8.77	& 8.75	&0.5 \\
2.88	&6,358	&0.167&	7.63&	0.408&2.00	&8.89	&8.98 &	8.93&	1.1 \\
3.08&	7,257&	0.169	&7.57&	0.450&	1.64&	8.78&	8.88&8.83&	1.2 \\
$U=30$ m/s	 \\
1.20	&6,430	&0.140&	8.45	&0.190	&6.08&	10.30	&10.72&	10.51&3.9 \\
1.80&8,588	&0.145&	8.41	&0.207	&5.63	&10.24&10.32	&10.28&0.8 \\
1.24	&10,997	&0.145&	8.44	&0.247&	4.31&	10.29&	10.32	&10.31&	0.4 \\
2.64&	14,208	&0.147&	8.39	&0.306&	2.91	&10.20	&10.20&	10.20	&0.1 \\
2.88	&16,584	&0.148	&8.38	&0.346&2.23	&10.19	&10.17	&10.18	&0.2 \\
3.08&	19,133	&0.145	&8.45	&0.388	&1.71	&10.31&	10.35&	10.33&	0.4 \\
\end{tabular}

\begin{tabular}{rrrrrrrrrr} \\
{\bf x, m} & ${\boldmath{Re_{\th}}}$ & ${\boldmath\a}$ &
{\bf A} & ${\boldmath\b}$ & {\bf B}  & ${\boldmath{\ln Re_1}}$ &
  ${\boldmath{\ln Re_2}}$ & ${\boldmath{\ln\RL}}$ & ${\boldmath{\D}}$\\

{\bf M.B. Jones data}\\
$U=10$ m/s	 \\
0.18	&855&0.144&	8.39&	0.20&	6.36	&10.21	&10.45	&10.33&	2.4 \\
0.40	&1,122	&0.144&	8.37&	0.176&	7.11	&10.17&	10.40	&10.29&	2.2 \\
0.60	&1,314	&0.146	&8.28	&0.168	&7.41&	10.01	&10.25&	10.13	&2.4 \\
0.80	&1,466	&0.148	&8.19	&0.166	&7.47	&9.86&10.11	&9.98&	2.5 \\
1.00	&1,616	&0.144&	8.38	&0.160&	7.68	&10.19	&10.44	&10.31&	2.5 \\
1.20	&1,745&	0.145&	8.35&	0.156	&7.84&10.13	&10.38&	10.25	&2.4 \\
1.40	&1,888	&0.142&8.44&	0.153	&7.99	&10.29	&10.55&	10.42&	2.5 \\
1.60	&2,039&	0.142	&8.45&0.150&	8.10&	10.28&10.53&	10.41&	2.4 \\
1.80&2,150&	0.143	&8.41&	0.148	&8.18	&10.23	&10.50	&10.36	&2.6 \\
2.00&	2,299&	0.141&	8.49&	0.144&	8.35&	10.37&	10.62&	10.50&2.4 \\
2.20	&2,411	&0.144&8.37	&---&	---	&10.17	&10.43&	10.30&	2.5 \\
2.40	&2,489	&0.139&	8.57&	--- &---	&10.52&	10.78&	10.65&	2.4 \\
2.60	&2,574&	0.145&	8.32	&--- &---	&10.08&	10.36&	10.22&	2.7 \\
2.80	&2,683	&0.142	&8.47&	--- &---	&10.34	&10.60&	10.47&	2.5 \\
2.92&	2,728&	0.145&	8.31&	--- &---&	10.06	&10.33&	10.19&	2.7 \\
3.04	&2,819	&0.149&	8.15&	--- &---&	9.79	&10.06	&9.92	&2.8 \\
3.16	&2,832	&0.147&8.24	&--- &---&9.94	&10.20&10.07&	2.6 \\
3.28	&2,946	&0.149&	8.14&	--- &---&	9.77&	10.05	&9.91&	2.8 \\
3.40	&2,987&0.142&	8.46	&--- &---	&10.32	&10.60	&10.46	&2.7
\\ 3.48	&3,026&0.145	&8.33	&--- &---&	10.11&	10.38&	10.24&	2.7 \\
3.54	&3,032&	0.146&	8.29&	--- &---	&10.03	&10.30	&10.16&	2.7 \\
3.58&	3,100&	0.146&8.27&	--- &---	&9.99&	10.28	&10.13	&2.9 \\
3.62	&3,029	&0.147	&8.20	&--- &---&	9.88&	10.20	&10.04&	3.2
\end{tabular}

\bs
For our subsequent analysis we will use the series corresponding to
$U=30$ m/s  of (2) and $U=10$ m/s of (4) for the following reasons: in spite of
a considerable variation in the usual parameter $Re_{\th}$, the effective
\Rn
$\RL$ obtained by the the procedure we introduced remains nearly constant and close,  for $U=30$ m/s (2), to a   
constant
$\ln\RL=10.3$,
and for $U=10$ m/s (4) to a constant $\ln Re_{\Lambda}=10.2$. The mean velocity
distribution  in bilogarithmic coordinates for both series is presented in
Figure 2. Thus, we are able to obtain, with some approximation,  
cross-sections
of the surfaces
$B(\RL,P)$, $\b(\RL,P)$. The results corresponding to $\ln\RL=10.3$
(adverse pressure gradient) are presented in Figures 3(a) and 3(b);
results corresponding to $\ln\RL=10.2$
(favourable pressure gradient) are presented in Figures 3(c) and 3(d).
Note that for large values of the favourable  pressure gradient we were
unable to reveal the second self-similar region. The situation is reminiscent of 
the disappearance of the second self-similar region in flows with an elevated
level of free-stream turbulence. We found such a situation previously (10) when we
processed the results of the remarkable experimental work of P.E.~Hancock and
P.~Bradshaw (14).

In the papers (1)-(4) the results concerning pressure were presented
through a coefficient
\[
C_p \ = \ \frac{p-p_\i}{\frac 12 \rho U^2}
\]
where $p_\i$ is a constant reference pressure. Therefore we calculated the
parameter $P$ using the relation $\p_x p=\frac 12 \rho U^2\p_x C_p$ where
the density $\rho$ cancelled out; the values of all the other parameters are available in (2),(4). The
values of the parameter
$P$ for $U=30$ m/s (2) and for $U=10$ m/s (4) are presented in Table 2.

\begin{center}{\bf Table 2}\\
\begin{tabular}{rrrrrrr}\\
{\bf I.Marusic}		\\
$Re_{\th}$ &	6,430&	8,588	&10,997&	14,208	&16,584& 
	19,133\\
$P*10^3$&	0	&1.75	&2.86&4.2	&5.79&7.04\\
$\ln Re_{\Lambda}$ &	10.5	&10.3	&10.3&	10.2	&10.2	&10.3\\
{\bf M.B.Jones	}\\
$Re_{\th}$ &	855&	1,122	&1,314&	1,466&	1,616&	1,745\\
$-P*10^3$ & 1.8 & 2.36 & 2.69 & 2.78 & 2.76 & 2.8\\
$\ln Re_{\Lambda}$ &		10.3&	10.3&	10.1	&10.0&10.3	&10.2
\end{tabular}\end{center}

\bs
Eliminating the parameter $P$ from relations [6], we obtain:
$$
B \ = \ B(\RL,\b) \ . \eqno{[7]}
$$
This relation is presented in Figure 4 in the form of a dependence of
$B$ on $\frac 1\b$. We see that this dependence is close to linear:
$$
B \ = \ \frac{1.75}{\b} - 2.80  \eqno{[8]}
$$
for the data by  Maru\v{s}i\'c (2) ({\it adverse} pressure gradient)
and
$$
B \ = \ \frac{1}{\b} + 1.43  \eqno{[9]}
$$
for the data by  Jones (4) ({\it favourable} pressure gradient)

For layer I there is also a linear relation between the
coefficients $A$ and $\frac 1\a$, but contrary to $B=B(\frac 1\b)$
this relation is universal.  The coefficients in the relation
$B=B(\frac 1\b)$ should in principle depend on $\RL$.

\section{Conclusion}

A new similarity parameter is obtained for the flow in the
upper self-similar region of a developed non-zero-\pg \t t \bl . Comparison with
\ep data for nearly  constant effective Reynolds numbers revealed simple  (close 
to linear) Reynolds number-dependent
relations between the parameters of the scaling law for the mean velocity
distributions in the upper \ss  layer.

The investigation performed in the present Note and the papers
(9)-(12) demonstrated that the  Reynolds number-dependent scaling law for
the velocity distribution across the shear flow obtained initially for
flows in pipes is valid (with the same values of the constants) for the
developed turbulent boundary layer flows. This allows us to expect that
this scaling law reflects a universal property of all developed shear
flows. The  Reynolds number entering the law cannot be selected
arbitrarily, for example, as $Re_{\theta}$: it is uniquely determined by
the flow itself. The simple  procedure for the determination of the
appropriate  Reynolds number, which we proposed earlier, has been further
validated in the present Note.

We expect that the same approach will work for more complicated flows:
mixing layers, jets and wall jets. However, the delicate task of
investigating such flows requires high quality experimental data which are
still lacking.

The concepts of incomplete similarity and vanishing viscosity asymptotics
which we used for shear flows lead to plausible results for the local
strucxture of developed turbulent flows. Here, however, high quality
experimental data are very rare, specially for the higher order structure
functions, where we have conjectured that divergences may occur.

\bs\bs{\bf Acknowledgement.} The authors express their gratitude
to Professor Ivan Maru\v{s}i\'c for clarification of experimental results.
The work was supported by the Applied Mathematics subprogram of the
U.S.~Department of Energy under contract DE--AC03--76--SF00098.

\bs\bs\section*{References}
\setlength{\baselineskip}{14pt}

\begin{enumerate}
\item Maru\v{s}i\'c, Ivan, and Perry, A.E. (1995).
{\it J. Fluid Mech.} {\bf 298}, 389--407.
\item Maru\v{s}i\'c, I. (1991). The structure of zero and
adverse-pressure gradient \t t boundary layers. PhD thesis,
University of Melbourne, Australia,\\
(http:/\!/www.mame.mu.oz.au/\!$\sim$\!ivan/).
\item Jones, M.B., Maru\v{s}i\'c, Ivan, and Perry, A.E. (2001).
{\it J. Fluid Mech.} {\bf 428}, 1--27.
\item Jones, M.B. (1998). Evolution and structure of sink-flow \t t boundary
layers. PhD thesis,
University of Melbourne, Australia,
(http:/\!/www.mame.mu.oz.au/\!$\sim$\!mbjones/).
\item Clauser, D.E. (1954). {\it J. Aero. Sci.} {\bf 21}, 91--108.
\item Coles, D.E. (1956). {\it J. Fluid Mech.} {\bf 1}, 1--51.
\item Prandtl, L. (1932). {\it Ergebn. Aerodyn. Versuchanstalt
Gottingen} {\bf 4}, 18--29.
\item von Karman, Th. (1930). In: C.W.Oseen and W.Weibull (eds.),
{\it Proc. 3rd Int. Cong. Appl. Mech.}  AB Sveriges Litografiska
Tryckenier, Stockholm. Vol. I, 85--93.
\item  Barenblatt, G.I., Chorin, A.J., Hald, O.H., and Prostokishin,
V.M. (1997). {\it Proc. Natl. Acad. Sci. USA} {\bf 94}, 7817--7819.
\item  Barenblatt, G.I., Chorin, A.J.,  and Prostokishin,
V.M. (2000). {\it J. Fluid Mech.} {\bf 410}, 263--283.
\item  Barenblatt, G.I., Chorin, A.J.,  and Prostokishin,
V.M. (2000). Ctr for Pure and Applied Mathematics, University of
California, Berkeley, California, report CPAM--777.
\item  Barenblatt, G.I., Chorin, A.J., and Prostokishin,
V.M. (2000). {\it Proc. Natl. Acad. Sci. USA} {\bf 97}, 3799--3802.
\item Van Dyke, M. (1980). {\it An Album of Fluid Motion}.
The Parabolic Press, Stanford, Calfornia.
\item Hancock, P.E., and Bradshaw, P. (1989).  {\it J. Fluid Mech.}
{\bf 205}, 45--76.
\end{enumerate}

\newpage\section*{Figure Captions}
\begin{description}
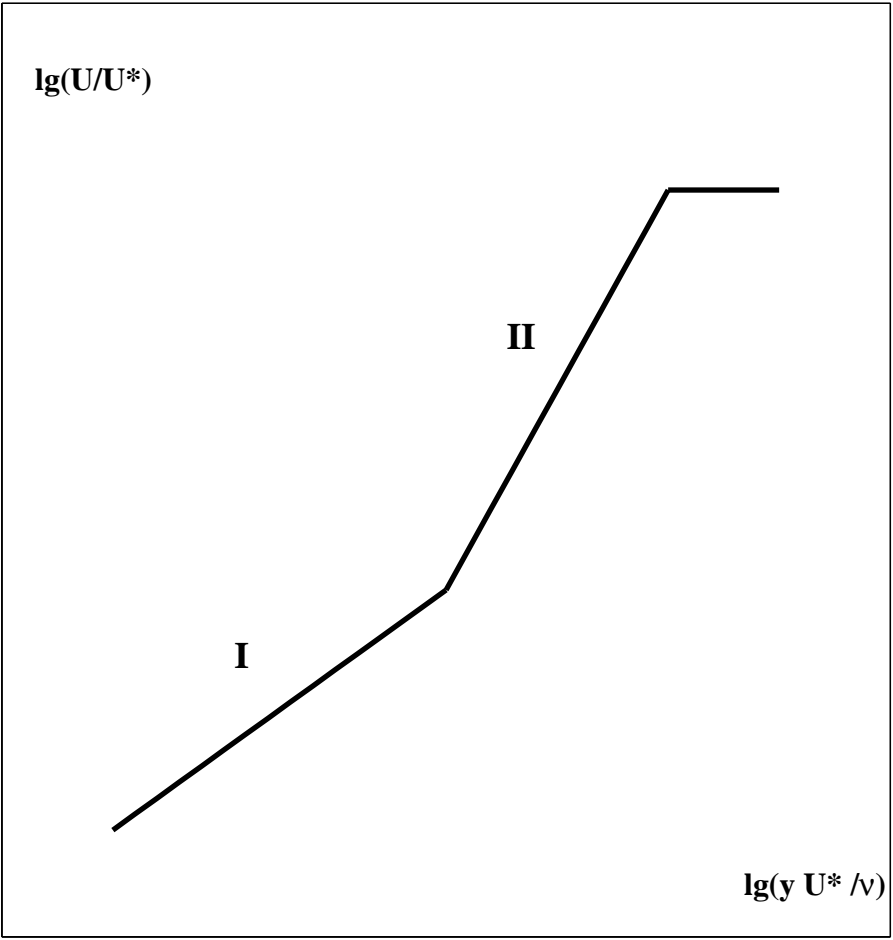
\item{Figure 1.} Schematic representation of the mean velocity profile
in developed \t t \bl \ in bilogarithmic coordinates
$\lg\frac{u}{u_*}$, \ $\lg\frac{u_*y}{\nu}$.

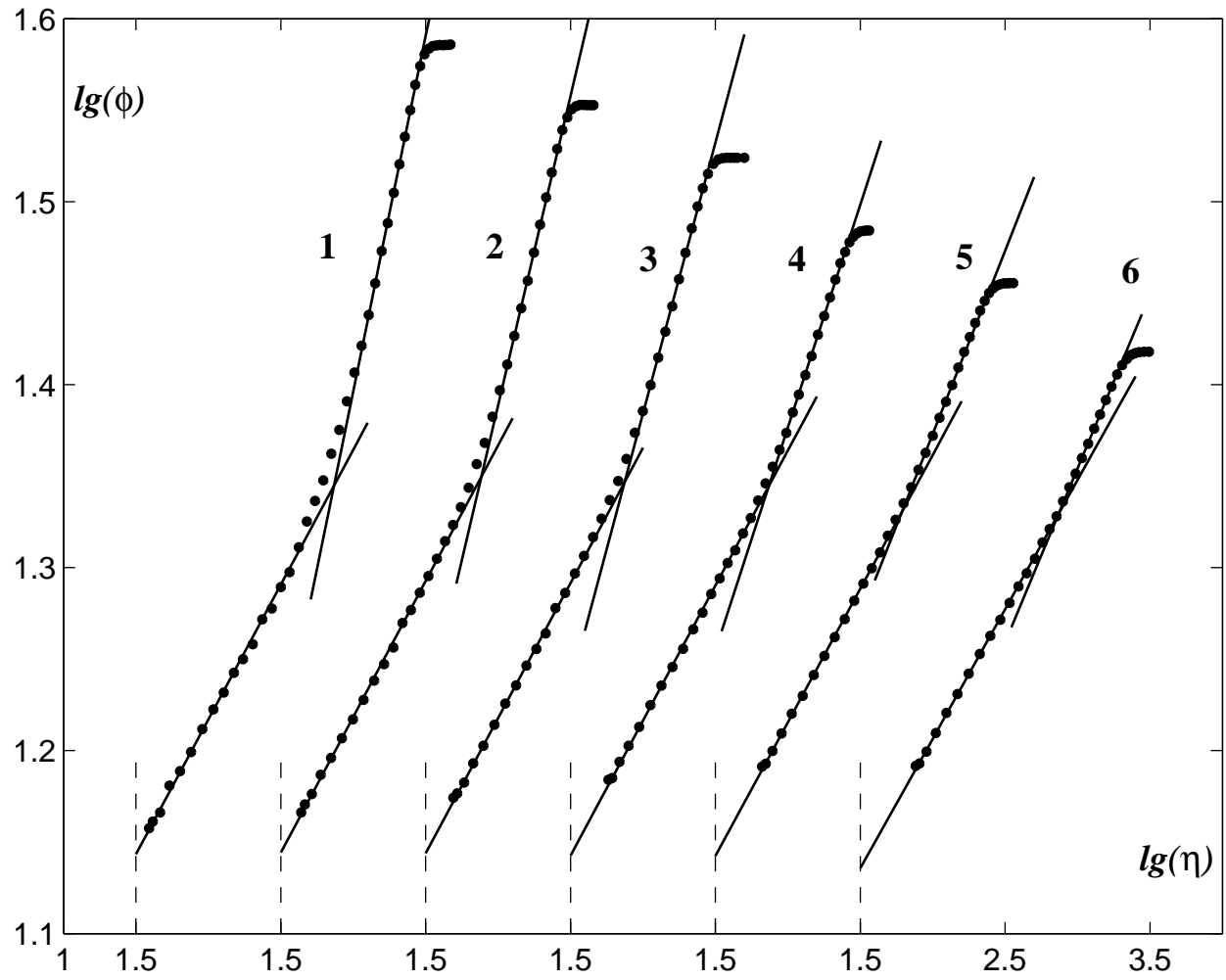
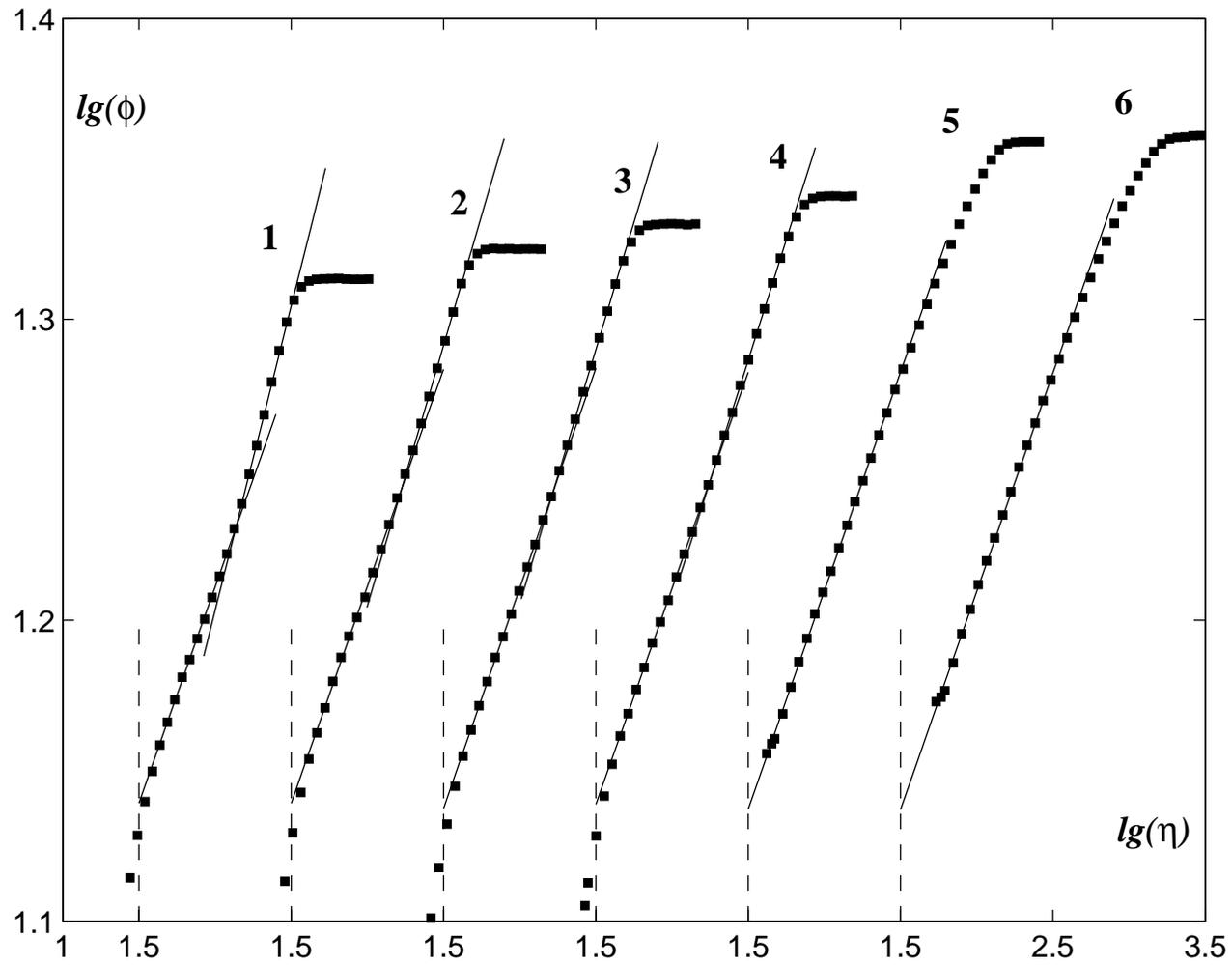
\item{Figure 2.} (a) The mean velocity profiles in  bilogarithmic coordinates
in the series of experiments of Maru\v{s}i\'c for $U=30$ m/s;
{\it adverse} pressure gradient
.\smallskip\\
\begin{tabular}{ll}
(1) $Re_{\th}= 19,133$ & \qquad (2) $Re_{\th}= 16,584$,\\
(3) $Re_{\th}= 14,208$ & \qquad (4) $Re_{\th}= 10,997$,\\
(5) $Re_{\th}= 8,588$ & \qquad (6) $Re_{\th}= 6,430$.\end{tabular}\smallskip\\
The `chevron' structure of the profiles is clearly seen
and regions I and II are clearly distinguishable.

(b) The mean velocity profiles in  bilogarithmic coordinates
in the series of experiments of Jones for $U=10$ m/s;
{\it favourable} pressure gradient
.\smallskip\\
\begin{tabular}{ll}
(1) $Re_{\th}= 855$ & \qquad (2) $Re_{\th}= 1,122$,\\
(3) $Re_{\th}= 1,314$ & \qquad (4) $Re_{\th}= 1,616$,\\
(5) $Re_{\th}= 2,728$ & \qquad (6) $Re_{\th}= 3,032$.\end{tabular}\\
The `chevron' structure of the profiles is clearly seen
for the curves (1)--(4), where $\b>\a$.

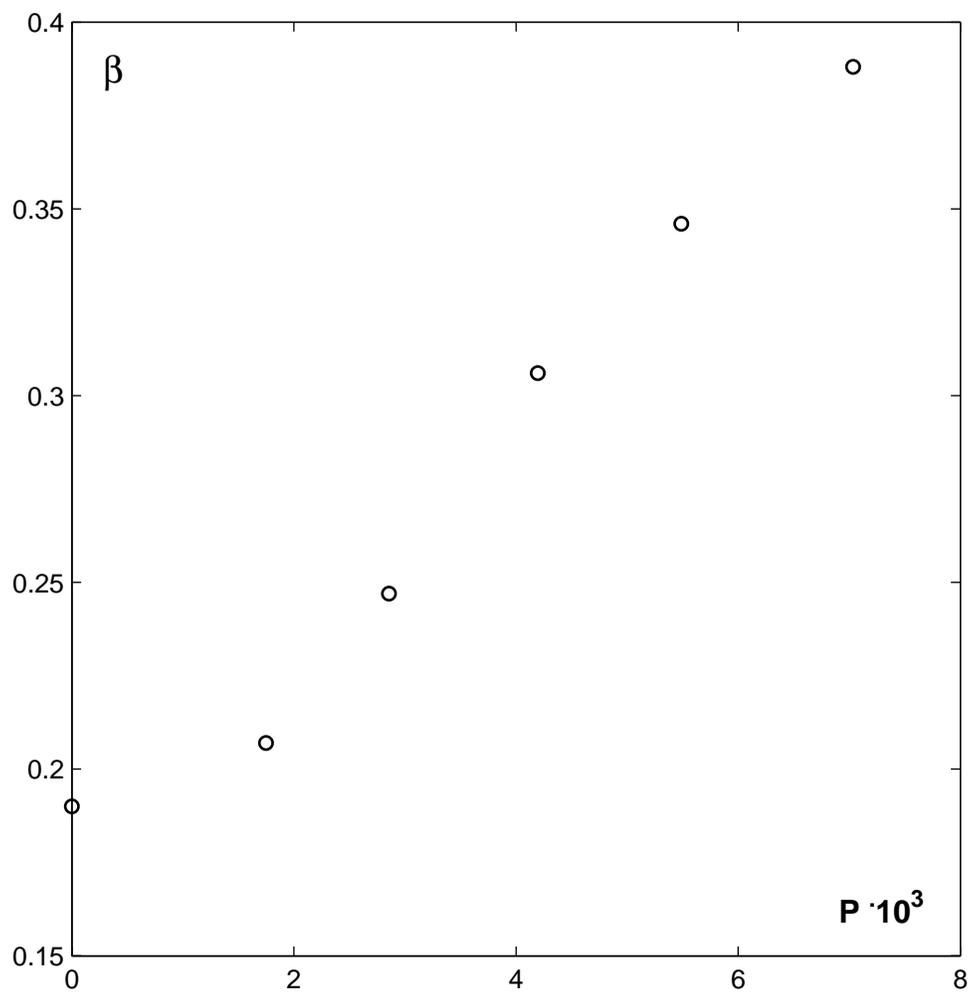
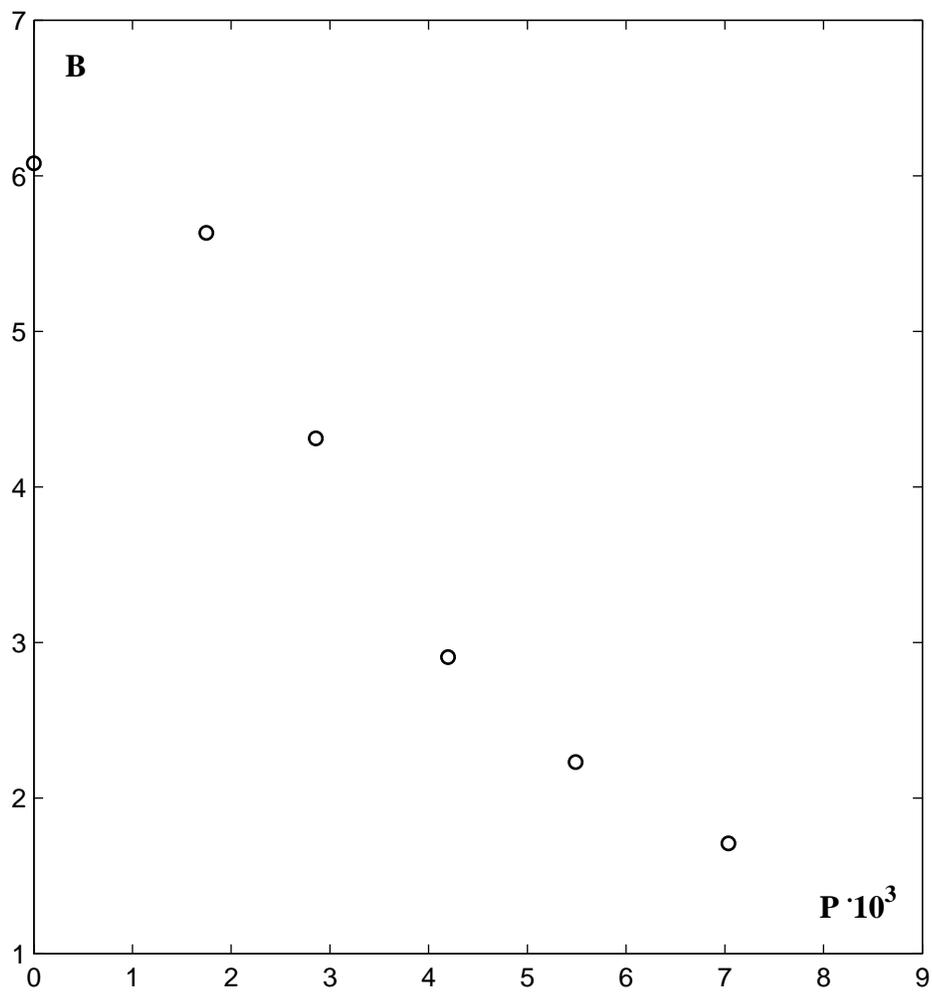
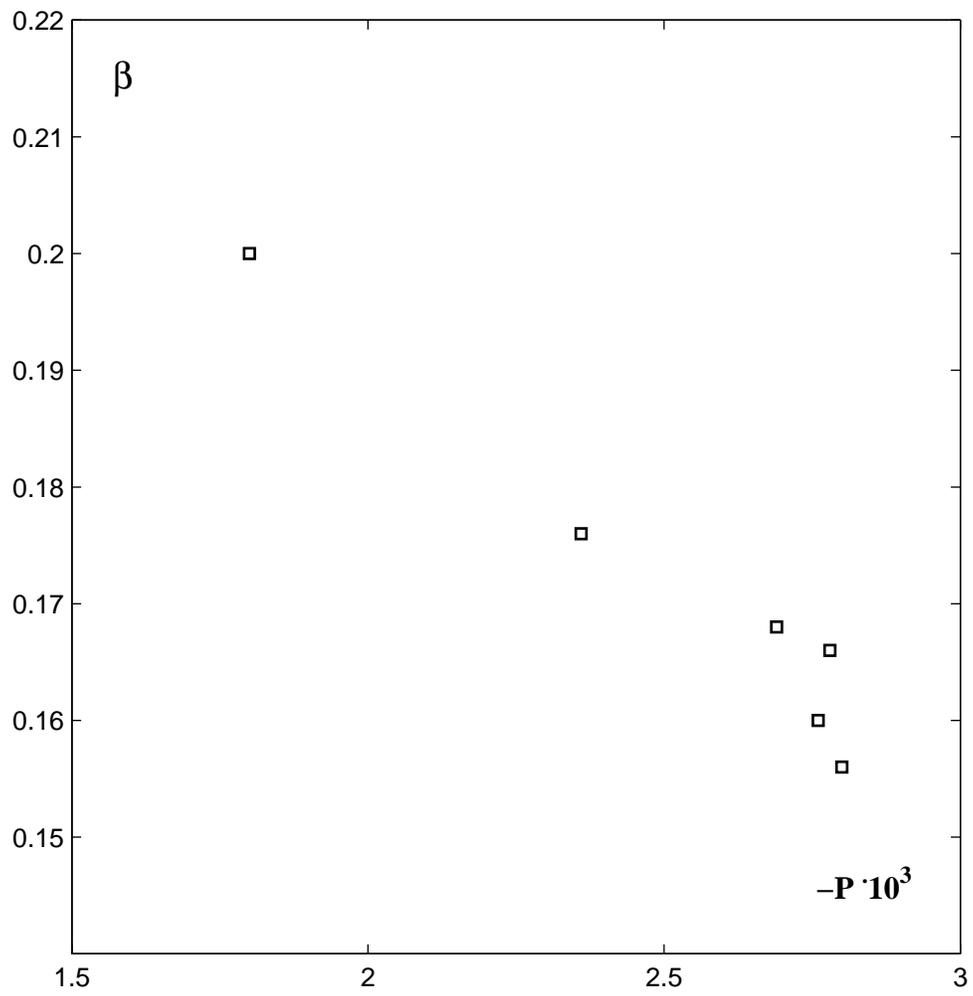
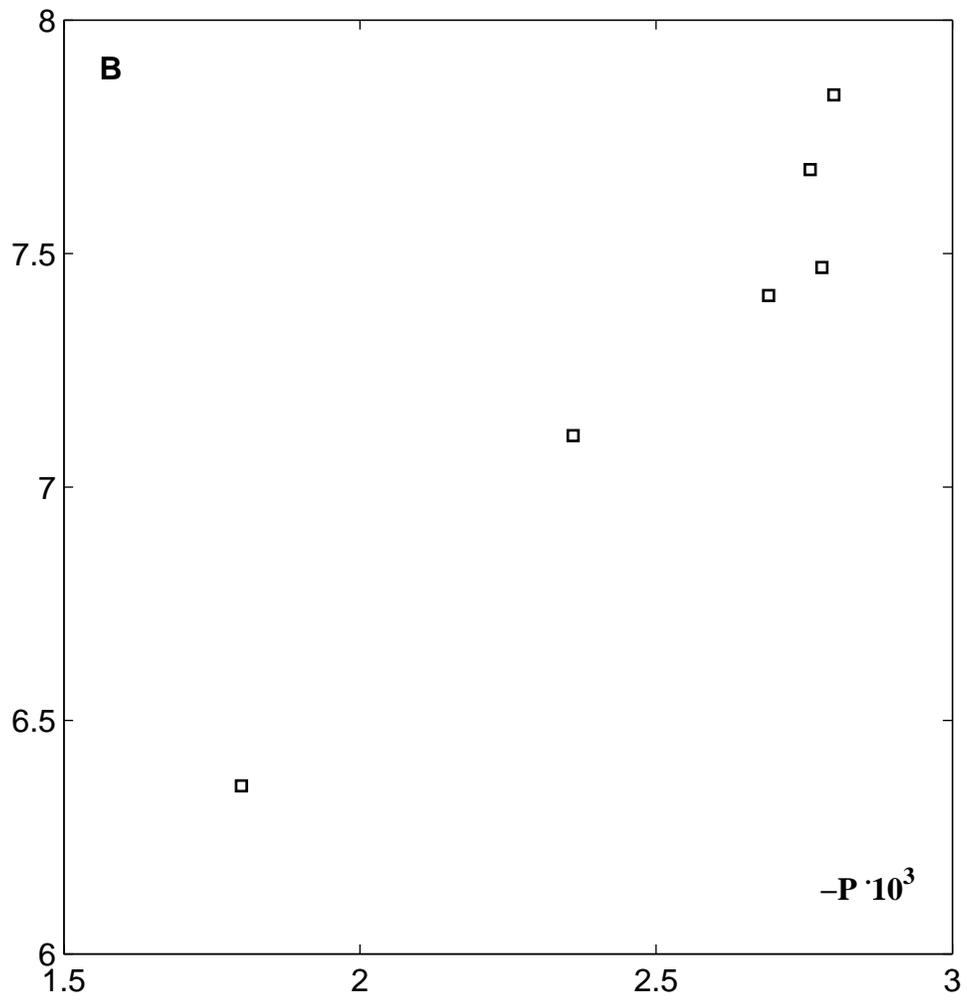
\item{Figure 3.} (a) Cross-section of the surface
$\b(\RL,P)$, for $\RL\cong 10.3$; (2).

(b) Cross-section of the surface
$B(\RL,P)$, for $\RL\cong 10.3$; (2).

(c) Cross-section of the surface
$\b(\RL,P)$, for $\RL\cong 10.2$; (4).

(d) Cross-section of the surface
$B(\RL,P)$, for $\RL\cong 10.2$; (4).

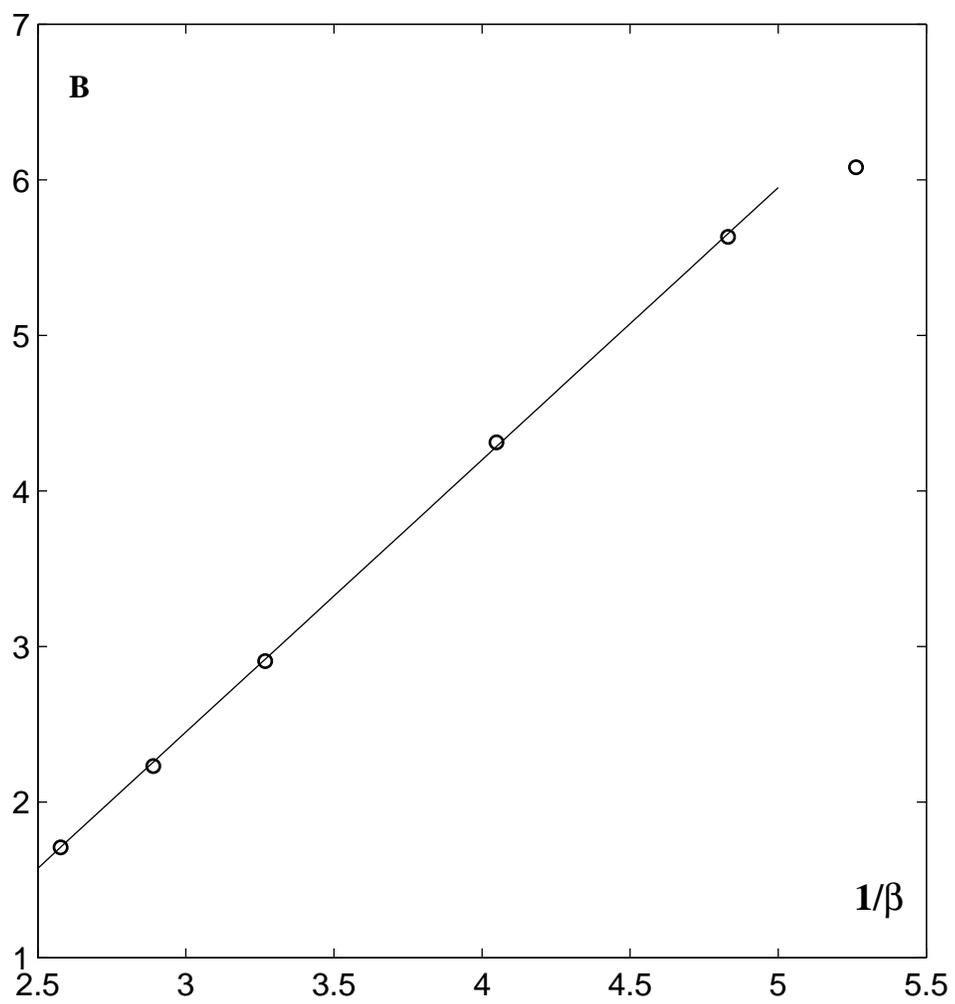
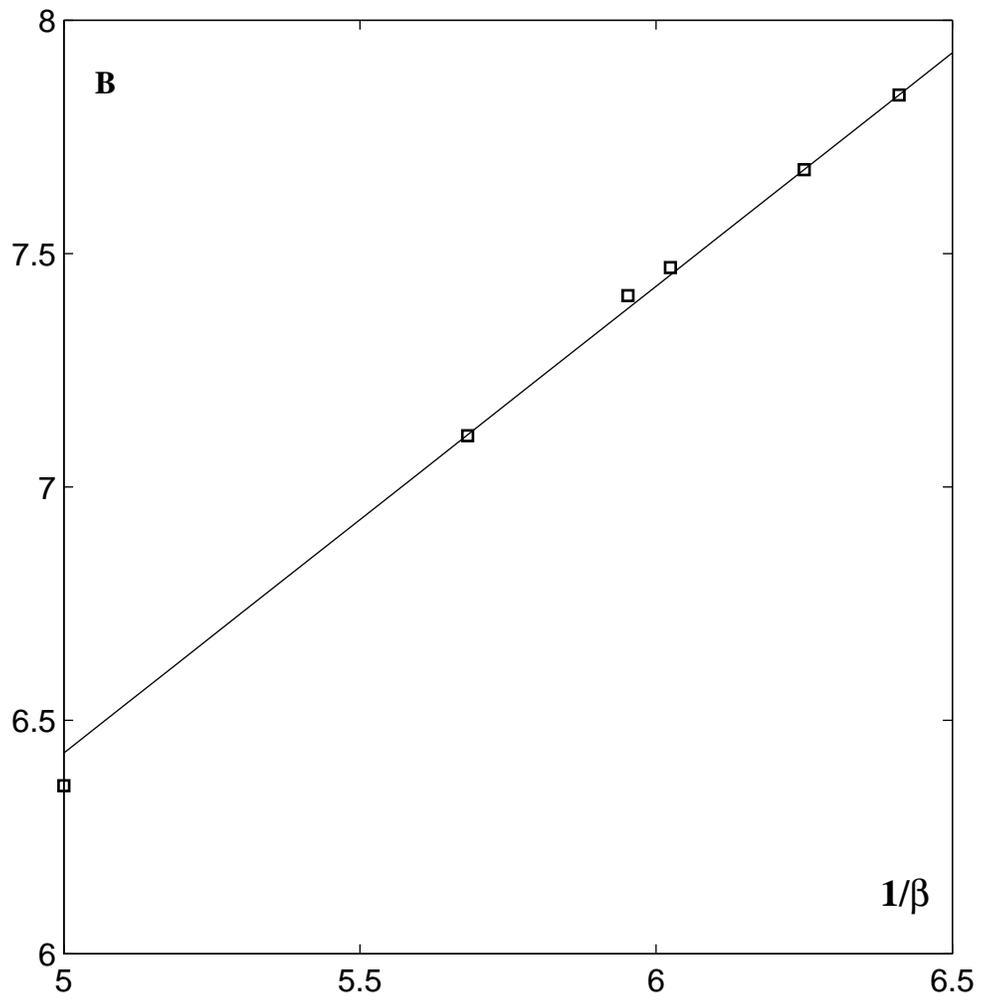
\item{Figure 4.} (a) The dependence $B(\frac 1\b)$ for $\RL\cong 10.3$;\\
the straight line corresponds to \ $1.75/\b -2.8$.

(b) The dependence $B(\frac 1\b)$ for $\RL\cong 10.2$;\\
the straight line corresponds to \ $1/\b +1.43$.

\end{description}

\end{document}